---------------------------------------------------------------------------
\documentstyle[12pt]{article}   \topmargin -48pt
\begin{document}

\begin{titlepage}
\begin{center}
{\large JOINT INSTITUTE FOR NUCLEAR RESEARCH\\
Laboratory of theoretical physics}\\
\end{center}
\vspace*{2cm}
\hspace*{10cm} JINR Communications\\
\hspace*{10cm} E2-92-440\\
\hspace*{10cm} October 1992\\
\vspace*{2cm}
\begin{center}
{\Large{\bf
QUASIPOTENTIAL IN THE FOURTH ORDER OF PERTURBATION THEORY.}}\\
{\large{\bf Unequal mass case.}}\\
\vspace{0.5cm}
{\bf N.A. Boikova$^1$, V.V. Dvoeglazov$^1$, Yu.N. Tyukhtyaev}
\footnote[1]{\it\noindent Permanent address:
Department of Theoretical and Nuclear Physics,
Saratov State University \\ and Scientific and Technical Center
for Control and Use of Physical Fields and Radiation,\\
Saratov 410071 RUSSIA}\\
{\bf R.N. Faustov}\footnote[2]{\it
\noindent Permanent address: Scientific Council
"Cybernetics",
Russian Academy of Sciences,\\ Moscow 117333 RUSSIA}
\end{center}

\vspace*{0.5cm}
\begin{abstract}
The quasipotential in the fourth order of perturbation theory is calculated
in the Coulomb gauge for the unequal mass particles. It could be
used for the future calculations of energy spectra in two-body
systems.
\end{abstract}
\vspace*{7cm}
\centerline{Dubna, 1992}
\end{titlepage}

At present the problem of an existence of the narrow resonances in ($e^+e^-$)-
and (pp)-systems \cite{a1,a2} requires the adequate theoretical description
{}~\cite{a3}-\cite{a6}. The method proposed in the above papers is based on the
quasipotential equation \cite{a7,a8} with the relativistic Coulomb potential
modified with taking into account the binding effects in two-particle
systems.

In this article the method of construction of the quasipotential by means of
the two-time Green function
$\widehat{G}(x_a,x_b,t,y_a,y_b,t')$ is applied to the investigation
of processes of the two-photon exchange.

The quasipotential
\begin{eqnarray}
\hat V&=&F^{-1}-(\widehat G^+)^{-1}\\
F&=&\widehat G^+_0=(2\pi)^3\delta(\vec p-\vec q)(E-\sqrt{\vec
p^2+m_1^2}-\sqrt{\vec p^2+m_2^2})^{-1}
\end{eqnarray}
was used earlier for the calculation of  corrections to the Fermi
energy of the hyperfine splitting and for the analysis of the fine structure
energy in  hydrogen-like atoms \cite{a9}-\cite{a11}.

Recently the papers \cite{a12,a13} \footnote[1]{The results of
these papers are different for the positronium from each other:
\begin{eqnarray}
\Delta E_{\cite{a12}}&=&\frac{1}{96}m\alpha^6log\alpha^{-1}(3+
5\vec \sigma_1\vec \sigma_2)\frac{\delta_{l0}}{n^3}   \\
\Delta E_{\cite{a13}}&=&\frac{5}{96}m\alpha^6log\alpha^{-1}(3+\vec \sigma_1\vec
\sigma_2)\frac{\delta_{l0}}{n^3}
\end{eqnarray}}
devoted to the calculation of the  $O(\alpha^6 log\alpha)$ corrections in the
two-body quantum electrodynamic systems have been appeared. The planned
experiments will produce the question of eliminating the discrepancy in the
results of calculation of the
$O(\alpha^6 log\alpha)$ corrections between \cite{a12} and \cite{a13}.

The expressions for the quasipotential in the fourth order  of  perturbation
theory (unequal mass case), which are presented in this paper,
could be used for  verifying the results \cite{a12,a13} and solving the
problem.
The indices C and T describe the successive exchange of the Coulomb photons and
the transverse photons, respectively \footnote[2]{ The Coulomb gauge is used. }
The index "it" marks the expressions corresponding to the iteration diagrams.
The index "$\times$" marks the expressions corresponding
to the cross diagrams ({\bf Fig.}1b,1e,1f,1h ).

\begin{eqnarray}
\hat V_{CC}( \vec p, \vec q; E) &=& -\frac{2\alpha^2}{\pi} u_{1}^{*}(\vec p)
u_{2}^{*}(-\vec p)\int\frac{d\vec k}{k_{p}^{2}k_{q}^{2}}
\left\{ \frac{\Lambda^{+}_{1}(\vec k)\Lambda^{+}_{2}(-\vec k)}
{\epsilon_{1k}+\epsilon_{2k}-E}+\right.\nonumber\\
&+&\left.\frac{\Lambda^{-}_{1}(\vec k)\Lambda^{-}_{2}(-\vec k)}
{\epsilon_{1k}+\epsilon_{2k}+E} \right\} u_{1}(\vec q)u_{2}
(-\vec q)
\end{eqnarray}

\begin{eqnarray}
\hat V^{\times}_{CC}( \vec p, \vec q; E) &=& \frac{2\alpha^2}{\pi}
u_{1}^{*}(\vec p) u_{2}^{*}(-\vec p)\int\frac{d\vec k}{k_{p}^{2} k_{q}^{2}}
\left\{ \frac{\Lambda^{+}_{1}(\vec k)\Lambda^{-}_{2}(\vec k-\vec p-\vec
q)}{\epsilon_{2p}+\epsilon_{1k}+\epsilon_{2kpq}+\epsilon_{2q}-E}+
\right.\nonumber\\
&+&\left.\frac{\Lambda^{-}_{1}(\vec k)\Lambda^{+}_{2}(\vec k-\vec p-\vec
q)}{\epsilon_{1p}+\epsilon_{1k}+\epsilon_{2kpq}+\epsilon_{1q}-E}\right\}
u_{1}(\vec q)u_{2}(-\vec q)
\end{eqnarray}
\newpage

\begin{eqnarray}
\lefteqn{\hat V_{CT}( \vec p, \vec q ; E) = \frac{\alpha^2}{\pi} u_{1}^{*}(\vec
p) u_{2}^{*}(-\vec p)\int\frac{d\vec k}{k_{p}^{2}k_{q}}
\left\{ \frac{\Lambda^{+}_{1}(\vec k)\Lambda^{+}_{2}(-\vec
k)}{\epsilon_{1k}+\epsilon_{2k}-E}(\frac{1}{k_q+\epsilon_{2k}+
\epsilon_{1q}-E}+\right .}\nonumber\\
&+&\left.\frac{1}{k_q+\epsilon_{1k}+\epsilon_{2q}-E})+
\frac{\Lambda^{-}_{1}(\vec k)\Lambda^{-}_{2}(-\vec
k)}{\epsilon_{1k}+\epsilon_{2k}+E}(\frac{1}{k_q+\epsilon_{1k}+
\epsilon_{1q}}+\frac{1}{k_q+\epsilon_{2k}+\epsilon_{2q}})-\right.
\nonumber\\
&-&\left.\frac{\Lambda^{-}_{1}(\vec k)\Lambda^{+}_{2}(-\vec
k)}{(k_q+\epsilon_{2k}+\epsilon_{1q}-E)(k_q+\epsilon_{1k}+
\epsilon_{1q})}-\frac{\Lambda^{+}_{1}(\vec k)\Lambda^{-}_{2}(-\vec
k)}{(k_q+\epsilon_{1k}+\epsilon_{2q}-E)(k_q+\epsilon_{2k}+
\epsilon_{2q})} \right\}\times\nonumber\\
&\times&\Gamma_{12}(\vec k-\vec q)u_{1}(\vec q)u_{2}(-\vec q)
\end{eqnarray}

\begin{eqnarray}
\lefteqn{\hat V^{\times}_{CT}( \vec p, \vec q; E)=\frac{\alpha^2}{\pi}
u_{1}^{*}(\vec p) u_{2}^{*}(-\vec p)\int\frac{d\vec k}{k_{p}^{2}k_q}
\left\{ \frac{\Lambda^{+}_{1}(\vec k)\Gamma_{12}(\vec k-\vec
q)\Lambda^{+}_{2}(\vec k-\vec p-\vec
q)}{(k_q+\epsilon_{1p}+\epsilon_{2kpq}-E)(k_q+\epsilon_{1k}+
\epsilon_{2q}-E)}+\right.}\nonumber\\
&+&\left.\frac{\Lambda^{-}_{1}(\vec k)\Gamma_{12}(\vec k-\vec
q)\Lambda^{-}_{2}(\vec k-\vec p-\vec
q)}{(k_q+\epsilon_{2p}+\epsilon_{2kpq})(k_q+\epsilon_{1k}+\epsilon_{1q})}-
\frac{\Lambda^{+}_{1}(\vec k)\Gamma_{12}(\vec k-\vec q)\Lambda^{-}_{2}(\vec
k-\vec p-\vec q)}{\epsilon_{2p}+\epsilon_{1k}+\epsilon_{2kpq}+\epsilon_{2q}-E}
\times\right.\nonumber\\
&\times&\left.(\frac{1}{k_q+\epsilon_{2p}+\epsilon_{2kpq}}+
\frac{1}{k_q+\epsilon_{1k}+\epsilon_{2q}-E})-
\frac{\Lambda^{-}_{1}(\vec k)\Gamma_{12}(\vec k-\vec q)\Lambda^{+}_{2}(\vec
k-\vec p-\vec q)}{\epsilon_{1p}+\epsilon_{1k}+\epsilon_{2kpq}+\epsilon_{1q}
-E}\times\right.\nonumber\\
&\times&\left.(\frac{1}{k_q+\epsilon_{1p}+\epsilon_{2kpq}-E}+
\frac{1}{k_q+\epsilon_{1k}+\epsilon_{1q}})\right\} u_{1}(\vec q)u_{2}(-\vec q)
\end{eqnarray}

\begin{eqnarray}
\lefteqn{\hat V_{TT}( \vec p, \vec q ; E) = -\frac{\alpha^2}{2\pi}
u_{1}^{*}(\vec p) u_{2}^{*}(-\vec p)\int\frac{d\vec k}{k_{p}
k_{q}}\Gamma_{12}(\vec p-\vec k)\left\{
\frac{\Lambda^{+}_{1}(\vec k)\Lambda^{+}_{2}(-\vec
k)}{\epsilon_{1k}+\epsilon_{2k}-E}\times\right.}\nonumber\\
&\times&\left.\left [
\frac{1}{k_p+k_q+\epsilon_{1p}+\epsilon_{2q}-E}(\frac{1}{k_p+
\epsilon_{1p}+\epsilon_{2k}-E}+\frac{1}{k_q+\epsilon_{1k}+
\epsilon_{2q}-E})+\right.\right.\nonumber\\
&+&\left.\left.\frac{1}{k_p+k_q+\epsilon_{2p}+\epsilon_{1q}-E}
(\frac{1}{k_p+\epsilon_{2p}+\epsilon_{1k}-E}+\frac{1}{k_q+
\epsilon_{2k}+\epsilon_{1q}-E})+\right.\right.\nonumber\\
&+&\left.\left.\frac{1}{(k_p+\epsilon_{2p}+\epsilon_{1k}-E)(k_q+
\epsilon_{1k}+\epsilon_{2q}-E)}+\frac{1}{(k_p+\epsilon_{1p}+
\epsilon_{2k}-E)(k_q+\epsilon_{2k}+\epsilon_{1q}-E)}\right ]
\right.+\nonumber\\
&+&\left.\frac{\Lambda^{-}_{1}(\vec k)\Lambda^{-}_{2}(-\vec
k)}{\epsilon_{1k}+\epsilon_{2k}+E}\left [
\frac{1}{k_p+k_q+\epsilon_{1p}+\epsilon_{2q}-E}(\frac{1}{k_p+
\epsilon_{1p}+\epsilon_{2k}-E_1+E_2}+\right.\right.\nonumber\\
&+&\left.\left.\frac{1}{k_q+\epsilon_{1k}+\epsilon_{2q}+E_1-E_2})+\frac{1}{k_p+k_q+\epsilon_{2p}+\epsilon_{1q}-E}\times\right.\right.\nonumber\\
&\times&\left.\left.(\frac{1}{k_p+\epsilon_{2p}+\epsilon_{1k}+
E_1-E_2}+\frac{1}{k_q+\epsilon_{2k}+\epsilon_{1q}-E_1+E_2})+
\right.\right.\nonumber\\
&+&\left.\left.\frac{1}{(k_p+\epsilon_{2p}+\epsilon_{1k}+
E_1-E_2)(k_q+\epsilon_{1k}+\epsilon_{2q}+E_1-E_2)}+
\right.\right.\nonumber\\
&+&\left.\left.\frac{1}{(k_p+\epsilon_{1p}+\epsilon_{2k}-E)(k_q+
\epsilon_{2k}+\epsilon_{1q}-E)}\right ]+\frac{\Lambda^{-}_{1}
(\vec k)\Lambda^{+}_{2}(-\vec
k)}{-\epsilon_{1k}+\epsilon_{2k}-E}\times\right.\nonumber \\
&\times&\left.\left [
\frac{1}{k_p+k_q+\epsilon_{1p}+\epsilon_{2q}-E}(\frac{1}{k_p+
\epsilon_{1p}+\epsilon_{2k}-E}-\frac{1}{k_p+\epsilon_{1p}+
\epsilon_{1k}})+\right.\right.\nonumber\\
&+&\left.\left.\frac{1}{k_p+k_q+\epsilon_{2p}+\epsilon_{1q}-E}
(\frac{1}{k_q+\epsilon_{2k}+\epsilon_{1q}-E}-\frac{1}{k_q+
\epsilon_{1k}+\epsilon_{1q}})+\right.\right.\nonumber\\
&+&\left.\left.\frac{1}{(k_p+\epsilon_{1p}+\epsilon_{2k}-E)(k_q+
\epsilon_{2k}+\epsilon_{1q}-E)}-\frac{1}{(k_p+\epsilon_{1p}+
\epsilon_{1k})(k_q+\epsilon_{1k}+\epsilon_{1q})}\right ] +\right.\nonumber\\
&+&\left.\frac{\Lambda^{+}_{1}(\vec k)\Lambda^{-}_{2}(-\vec
k)}{\epsilon_{1k}-\epsilon_{2k}-E}\left [
\frac{1}{k_p+k_q+\epsilon_{1p}+\epsilon_{2q}-E}(\frac{1}{k_q+
\epsilon_{1k}+\epsilon_{2q}-E}-\frac{1}{k_q+\epsilon_{2k}+
\epsilon_{2q}})+\right.\right.\nonumber\\
&+&\left.\left.\frac{1}{k_p+k_q+\epsilon_{2p}+\epsilon_{1q}-E}
(\frac{1}{k_p+\epsilon_{2p}+\epsilon_{1k}-E}-\frac{1}{k_p+
\epsilon_{2p}+\epsilon_{2k}})+\right.\right.\nonumber\\
&+&\left.\left.\frac{1}{(k_p+\epsilon_{2p}+\epsilon_{1k}-E)
(k_{q}+\epsilon_{1k}+\epsilon_{2q}-E)}-\frac{1}{(k_p+
\epsilon_{2p}+\epsilon_{2k})(k_q+\epsilon_{2k}+\epsilon_{2q})}
\right ] \right\}\times\nonumber\\
&\times&\Gamma_{12}(\vec k-\vec q)u_{1}(\vec q)u_{2}(-\vec q)
\end{eqnarray}

\begin{eqnarray}
\lefteqn{\hat V^{\times}_{TT}( \vec p, \vec q ; E) = -\frac{\alpha^2}{2\pi}
u_{1}^{*}(\vec p) u_{2}^{*}(-\vec p)\int\frac{d\vec k}{k_p
k_q}\alpha_{1i}(\delta_{ij}-\frac{(\vec p-
\vec k)_i(\vec p-\vec k)_j}{(\vec p-\vec k)^2})\times}\nonumber\\
&\times&\left\{ \Lambda^{+}_{1}(\vec k)\Gamma_{12}(\vec k-
\vec q)\Lambda^{+}_{2}(\vec k-\vec p-\vec q)\left [
\frac{1}{k_p+k_q+\epsilon_{1k}+\epsilon_{2kpq}-E}
(\frac{1}{k_p+\epsilon_{2p}+\epsilon_{1k}-E}+\right.\right.
\nonumber\\
&+&\left.\left.\frac{1}{k_q+\epsilon_{1p}+\epsilon_{2kpq}-E})
(\frac{1}{k_p+\epsilon_{2kpq}+\epsilon_{1q}-E}+\frac{1}{k_q+
\epsilon_{1k}+\epsilon_{2q}-E})+\right.\right.\nonumber\\
&+&\left.\left.\frac{1}{(k_p+k_q+\epsilon_{1p}+\epsilon_{2q}-E)
(k_q+\epsilon_{1p}+\epsilon_{2kpq}-E)(k_q+\epsilon_{1k}+
\epsilon_{2q}-E)}+\right.\right.\nonumber\\
&+&\left.\left.\frac{1}{(k_p+k_q+\epsilon_{2p}+\epsilon_{1q}-E)
(k_p+\epsilon_{2p}+\epsilon_{1k}-E)(k_p+\epsilon_{2kpq}+
\epsilon_{1q}-E)}\right ]+\right.\nonumber\\
&+&\left.\Lambda^{-}_{1}(\vec k)\Gamma_{12}(\vec k-\vec q)\Lambda^{-}_{2}(\vec
k-\vec p-\vec q)\left [ \frac{1}{k_p+k_q+\epsilon_{2kpq}+\epsilon_{1k}+E}
(\frac{1}{k_p+\epsilon_{1p}+\epsilon_{1k}}+\right.\right.
\nonumber\\
&+&\left.\left.\frac{1}{k_q+\epsilon_{2p}+\epsilon_{2kpq}})
(\frac{1}{k_q+\epsilon_{1k}+\epsilon_{1q}}+\frac{1}{k_p+
\epsilon_{2kpq}+\epsilon_{2q}})+\right.\right.\nonumber\\
&+&\left.\left.\frac{1}{(k_p+k_q+\epsilon_{1p}+\epsilon_{2q}-E)
(k_p+\epsilon_{1p}+\epsilon_{1k})(k_p+\epsilon_{2kpq}+
\epsilon_{2q})}+\right.\right.\nonumber\\
&+&\left.\left.\frac{1}{(k_p+k_q+\epsilon_{2p}+\epsilon_{1q}-E)
(k_q+\epsilon_{2p}+\epsilon_{2kpq})(k_q+\epsilon_{1k}+
\epsilon_{1q})}\right ]-\right.\nonumber\\
&-&\left.\Lambda^{+}_{1}(\vec k)\Gamma_{12}(\vec k-\vec q)\Lambda^{-}_{2}(\vec
k-\vec p-\vec q)\left [ \frac{1}{\epsilon_{2p}+\epsilon_{1k}+\epsilon_{2kpq}+
\epsilon_{2q}-E}(\frac{1}{k_p+\epsilon_{2kpq}+\epsilon_{2q}}+
\right.\right.\nonumber\\
&+&\left.\left.\frac{1}{k_p+\epsilon_{2p}+\epsilon_{1k}-E})
(\frac{1}{k_q+\epsilon_{2p}+\epsilon_{2kpq}}+\frac{1}{k_q+
\epsilon_{1k}+\epsilon_{2q}-E})+\right.\right.\nonumber\\
&+&\left.\left.\frac{1}{(k_p+k_q+\epsilon_{1p}+\epsilon_{2q}-E)
(k_p+\epsilon_{2kpq}+\epsilon_{2q})(k_q+\epsilon_{1k}+
\epsilon_{2q}-E)}+\right.\right.\nonumber\\
&+&\left.\left.\frac{1}{(k_p+k_q+\epsilon_{2p}+\epsilon_{1q}-E)
(k_p+\epsilon_{2p}+\epsilon_{1k}-E)(k_q+\epsilon_{2p}+
\epsilon_{2kpq})}\right ]-\right.\nonumber\\
&-&\left.\Lambda^{-}_{1}(\vec k)\Gamma_{12}(\vec k-\vec q)\Lambda^{+}_{2}(\vec
k-\vec p-\vec q)\left [\frac{1}{\epsilon_{1p}+\epsilon_{1k}+\epsilon_{2kpq}+
\epsilon_{1q}-E}(\frac{1}{k_p+\epsilon_{1p}+\epsilon_{1k}}+
\right.\right.\nonumber\\
&+&\left.\left.\frac{1}{k_p+\epsilon_{2kpq}+\epsilon_{1q}-E})
(\frac{1}{k_p+\epsilon_{1k}+\epsilon_{1q}}+\frac{1}{k_q+
\epsilon_{1p}+\epsilon_{2kpq}-E})+\right.\right.\nonumber\\
&+&\left.\left.\frac{1}{(k_p+k_q+\epsilon_{1p}+
\epsilon_{2q}-E)(k_p+\epsilon_{1p}+\epsilon_{1k})(k_q+
\epsilon_{1p}+\epsilon_{2kpq}-E)}+\right.\right.\nonumber\\
&+&\left.\left.\frac{1}{(k_p+k_q+\epsilon_{2p}+
\epsilon_{1q}-E)(k_p+\epsilon_{2kpq}+\epsilon_{1q}-E)(k_q+
\epsilon_{1k}+\epsilon_{1q})}\right ] \right\}\times\nonumber\\
&\times&\alpha_{2j} u_{1}(\vec q) u_{2}(-\vec q)
\end{eqnarray}

\begin{equation}
\hat V_{CC}^{it}(\vec p,\vec q; E) = -\frac{2\alpha^2}{\pi} u_{1}^{*}(\vec p)
u_{2}^{*}(-\vec p)\int\frac{d\vec k}{k_{p}^{2}
k_{q}^{2}}\frac{\Lambda^{+}_{1}(\vec k)\Lambda^{+}_{2}
(-\vec k)}{\epsilon_{1k}+\epsilon_{2k}-E}u_{1}(\vec q)u_{2}
(-\vec q)
\end{equation}
\begin{eqnarray}
\lefteqn{\hat V_{CT}^{it}(\vec p,\vec q; E) = \frac{\alpha^2}{\pi}
u_{1}^{*}(\vec p) u_{2}^{*}(-\vec p)\int\frac{d\vec k}{k_{p}^{2}
k_{q}}\frac{\Lambda^{+}_{1}(\vec k)\Lambda^{+}_{2}
(-\vec k)}{\epsilon_{1k}+\epsilon_{2k}-E}\times}\nonumber\\
&\times&(\frac{1}{k_q+\epsilon_{1k}+\epsilon_{2q}-E}+
\frac{1}{k_q+\epsilon_{2k}+\epsilon_{1q}-E})\Gamma_{12}
(\vec k-\vec q)u_{1}(\vec q)u_{2}(-\vec q)
\end{eqnarray}
\begin{eqnarray}
\lefteqn{\hat V_{TT}^{it}(\vec p,\vec q; E) =
-\frac{\alpha^2}{2\pi} u_{1}^{*}(\vec p) u_{2}^{*}(-\vec p)\int\frac{d\vec
k}{k_pk_q}\Gamma_{12}
(\vec p-\vec k)\frac{\Lambda^{+}_{1}(\vec k)\Lambda^{+}_{2}
(-\vec k)}{\epsilon_{1k}+\epsilon_{2k}-E}\times}\nonumber\\
&\times&(\frac{1}{k_p+\epsilon_{2p}+\epsilon_{1k}-E}+
\frac{1}{k_p+\epsilon_{1p}+\epsilon_{2k}-E})
(\frac{1}{k_q+\epsilon_{1k}+\epsilon_{2q}-E}+\nonumber\\
&+&\frac{1}{k_q+\epsilon_{2k}+\epsilon_{1q}-E})
\Gamma_{12}(\vec k-\vec q)u_{1}(\vec q)u_{2}(-\vec q)
\end{eqnarray}

Here $\Lambda^{\pm}$ are the projecting operators, $k_p=\mid\vec p-\vec k\mid$,
$k_q=\mid\vec k-\vec q\mid$,
$k_{pq}=\mid\vec k -\vec p-\vec q\mid$, $\Gamma_{12}
(\vec k)=\vec \alpha_1 \vec \alpha_2-(\vec \alpha_1\vec k)
(\vec \alpha_2\vec k)/ \vec k^2$,
$\vec \alpha_i=\gamma^0_i\vec \gamma_i$ and $\gamma_{\mu}=(\gamma^0$, $\vec
\gamma)$ --
Dirac's matrices.

The expressions corresponding to {\bf Fig.}1d,1e,1k  are obtained
from the expressions
(7), (8), (12) by means of the substitutions $p \Leftrightarrow q$
 and the displacement of the
projecting operators and $\Gamma_{12}$. The quasipotentials
for the diagrams with
a vertex part and a self-energy part are easily deduced from the one-photon
exchange
quasipotential presented e.g in \cite{a6,a9,a11,a14,a15}.

We are grateful to B. A. Arbuzov, E. E. Boos, V. G. Kadyshevsky
and V. I. Savrin for
useful discussions and to N. B. Skachkov for the hospitality.

\newpage
\unitlength=1.00mm
\special{em:linewidth 0.4pt}
\linethickness{0.4pt}
\begin{picture}(70.00,180.22)
\put(10.33,140.22){\line(1,0){39.33}}
\put(59.67,140.22){\line(1,0){40.33}}
\put(109.67,140.22){\line(1,0){40.00}}
\put(149.67,114.84){\line(-1,0){39.67}}
\put(49.33,114.84){\line(-1,0){38.67}}
\put(100.67,114.84){\line(-1,0){41.00}}
\put(20.00,70.11){\line(0,-1){3.01}}
\put(20.00,53.76){\line(0,-1){3.87}}
\put(20.00,48.17){\line(0,-1){3.44}}
\put(40.00,44.73){\line(0,1){3.87}}
\put(40.00,49.89){\line(0,1){3.87}}
\put(40.00,55.48){\line(0,1){3.87}}
\put(40.00,61.07){\line(0,1){3.87}}
\put(40.00,66.66){\line(0,1){3.44}}
\put(149.67,104.95){\line(-1,0){39.67}}
\put(100.00,104.95){\line(-1,0){40.00}}
\put(50.00,104.95){\line(-1,0){40.33}}
\put(60.00,80.00){\line(1,0){40.00}}
\put(110.00,80.00){\line(1,0){40.00}}
\put(10.00,70.11){\line(1,0){40.00}}
\put(60.00,70.11){\line(1,0){40.00}}
\put(100.00,45.16){\line(-1,0){40.00}}
\put(50.00,45.16){\line(-1,0){40.00}}
\put(80.00,21.51){\line(0,0){0.00}}
\put(80.00,21.51){\line(0,0){0.00}}
\put(80.00,21.51){\line(0,0){0.00}}
\put(80.00,21.51){\line(0,0){0.00}}
\put(80.00,21.51){\line(0,0){0.00}}
\put(9.33,80.00){\line(0,0){0.00}}
\put(59.33,9.89){\line(1,0){15.67}}
\put(85.00,9.89){\line(1,0){15.00}}
\put(100.00,34.84){\line(-1,0){14.67}}
\put(75.00,34.84){\line(-1,0){15.33}}
\put(80.00,21.51){\makebox(0,0)[cc]{F}}
\put(20.00,55.48){\line(0,1){3.87}}
\put(20.00,59.35){\line(0,1){0.43}}
\put(20.00,59.78){\line(0,1){0.43}}
\put(20.00,61.07){\line(0,1){3.87}}
\put(20.00,64.94){\line(0,1){0.43}}
\put(20.00,65.37){\line(0,0){0.00}}
\put(91.00,70.11){\line(-4,-5){3.33}}
\put(85.67,63.22){\line(-4,-5){3.00}}
\put(140.00,115.27){\line(0,1){4.30}}
\put(140.00,120.86){\line(0,1){4.30}}
\put(140.00,126.45){\line(0,1){3.01}}
\put(140.00,126.45){\line(0,1){4.30}}
\put(140.00,132.91){\line(0,1){3.87}}
\put(140.00,136.78){\line(0,1){1.72}}
\put(140.00,138.93){\line(0,-1){5.59}}
\put(140.00,133.34){\line(0,1){5.16}}
\put(120.00,138.50){\circle*{0.86}}
\put(120.00,134.20){\circle*{0.86}}
\put(120.00,129.89){\circle*{0.86}}
\put(120.00,125.16){\circle*{0.86}}
\put(120.00,120.86){\circle*{0.86}}
\put(120.00,116.99){\circle*{0.86}}
\put(92.00,32.26){\circle*{0.86}}
\put(92.00,27.96){\circle*{0.86}}
\put(92.00,24.09){\circle*{0.86}}
\put(92.00,19.78){\circle*{0.86}}
\put(92.00,15.48){\circle*{0.86}}
\put(92.00,11.18){\circle*{0.86}}
\put(26.66,-1.72){\line(0,-1){3.44}}
\put(117.66,34.84){\line(0,-1){4.30}}
\put(117.66,27.96){\line(0,-1){3.87}}
\put(117.66,21.51){\line(0,-1){3.01}}
\put(117.66,18.92){\line(0,-1){1.29}}
\put(117.66,15.48){\line(0,-1){3.44}}
\put(117.66,11.18){\line(0,-1){1.29}}
\put(79.67,5.16){\makebox(0,0)[cc]{k}}
\put(79.67,40.00){\makebox(0,0)[cc]{h}}
\put(30.33,40.00){\makebox(0,0)[cc]{g}}
\put(30.33,74.84){\makebox(0,0)[cc]{d}}
\put(79.33,74.84){\makebox(0,0)[cc]{e}}
\put(130.33,74.84){\makebox(0,0)[cc]{f}}
\put(130.33,110.11){\makebox(0,0)[cc]{c}}
\put(80.00,110.11){\makebox(0,0)[cc]{b}}
\put(30.33,110.11){\makebox(0,0)[cc]{ }}
\put(10.00,80.00){\line(1,0){40.00}}
\put(79.67,55.48){\line(-4,-5){3.00}}
\put(74.67,49.46){\line(-4,-5){3.33}}
\put(71.67,70.11){\line(4,-5){3.33}}
\put(76.33,64.08){\line(4,-5){3.67}}
\put(81.33,57.20){\line(4,-5){4.00}}
\put(87.00,49.89){\line(4,-5){3.67}}
\put(41.33,117.85){\circle{0.86}}
\put(41.33,121.29){\circle{0.86}}
\put(41.33,124.73){\circle{0.86}}
\put(41.33,128.60){\circle{0.86}}
\put(41.33,132.90){\circle*{0.86}}
\put(41.33,137.21){\circle*{0.86}}
\put(21.33,137.64){\circle*{0.86}}
\put(21.33,132.90){\circle*{0.86}}
\put(21.33,128.60){\circle*{0.86}}
\put(21.33,124.73){\circle*{0.86}}
\put(21.33,121.29){\circle*{0.86}}
\put(21.33,117.85){\circle*{0.86}}
\put(41.33,128.60){\circle*{0.86}}
\put(41.33,124.73){\circle*{0.86}}
\put(41.33,121.29){\circle*{0.00}}
\put(41.33,117.85){\circle*{0.86}}
\put(89.33,137.64){\circle{0.86}}
\put(87.33,135.06){\circle{0.67}}
\put(85.33,132.04){\circle{0.67}}
\put(82.67,129.03){\circle{0.86}}
\put(80.00,126.45){\circle{0.86}}
\put(77.33,123.44){\circle{0.86}}
\put(74.00,120.00){\circle{0.86}}
\put(71.00,116.13){\circle{0.86}}
\put(91.33,116.13){\circle*{0.86}}
\put(88.00,120.00){\circle*{0.86}}
\put(85.00,123.44){\circle*{0.86}}
\put(82.00,126.45){\circle*{0.86}}
\put(80.33,129.03){\circle*{0.86}}
\put(77.33,132.04){\circle*{0.86}}
\put(74.00,135.06){\circle*{0.86}}
\put(71.33,137.64){\circle*{0.86}}
\put(71.00,116.13){\circle*{0.86}}
\put(74.00,120.00){\circle*{0.86}}
\put(77.33,123.44){\circle*{0.86}}
\put(80.00,126.45){\circle*{0.86}}
\put(82.67,129.03){\circle*{0.86}}
\put(85.33,132.04){\circle*{0.86}}
\put(87.33,135.06){\circle*{0.86}}
\put(89.33,137.64){\circle*{0.86}}
\put(140.00,103.66){\circle*{0.86}}
\put(136.33,99.79){\circle*{0.86}}
\put(132.67,95.06){\circle*{0.86}}
\put(128.67,90.32){\circle*{0.86}}
\put(124.67,85.59){\circle*{0.86}}
\put(121.00,81.72){\circle*{0.86}}
\put(120.33,104.95){\line(3,-4){4.00}}
\put(126.33,96.78){\line(4,-5){3.00}}
\put(131.00,90.32){\line(4,-5){3.33}}
\put(135.67,84.30){\line(4,-5){3.33}}
\put(41.33,82.58){\circle{0.86}}
\put(41.33,86.02){\circle{0.86}}
\put(41.33,89.46){\circle{0.86}}
\put(41.33,93.33){\circle{0.86}}
\put(41.33,97.63){\circle*{0.86}}
\put(41.33,101.94){\circle*{0.86}}
\put(41.33,93.33){\circle*{0.86}}
\put(41.33,89.46){\circle*{0.86}}
\put(41.33,86.02){\circle*{0.00}}
\put(41.33,82.58){\circle*{0.86}}
\put(21.33,79.57){\line(0,1){3.87}}
\put(21.33,84.73){\line(0,1){3.87}}
\put(21.33,90.32){\line(0,1){3.87}}
\put(21.33,95.91){\line(0,1){3.87}}
\put(21.33,101.50){\line(0,1){3.44}}
\put(91.33,81.29){\circle*{0.86}}
\put(88.00,85.16){\circle*{0.86}}
\put(85.00,88.60){\circle*{0.86}}
\put(82.00,91.61){\circle*{0.86}}
\put(80.33,94.19){\circle*{0.86}}
\put(77.33,97.20){\circle*{0.86}}
\put(74.00,100.22){\circle*{0.86}}
\put(71.33,102.80){\circle*{0.86}}
\put(91.00,104.52){\line(-4,-5){3.33}}
\put(85.67,97.63){\line(-4,-5){3.00}}
\put(79.67,89.89){\line(-4,-5){3.00}}
\put(74.67,83.87){\line(-4,-5){3.33}}
\put(109.33,45.16){\line(1,0){15.67}}
\put(135.00,45.16){\line(1,0){15.00}}
\put(150.00,70.11){\line(-1,0){14.67}}
\put(125.00,70.11){\line(-1,0){15.33}}
\put(131.00,56.78){\makebox(0,0)[cc]{F}}
\put(142.00,67.53){\circle*{0.86}}
\put(142.00,63.23){\circle*{0.86}}
\put(142.00,59.36){\circle*{0.86}}
\put(142.00,55.05){\circle*{0.86}}
\put(142.00,50.75){\circle*{0.86}}
\put(142.00,46.45){\circle*{0.86}}
\put(117.67,46.45){\circle*{0.86}}
\put(117.67,50.75){\circle*{0.86}}
\put(117.67,55.05){\circle*{0.86}}
\put(117.67,59.36){\circle*{0.86}}
\put(117.67,63.23){\circle*{0.86}}
\put(117.67,67.53){\circle*{0.86}}
\put(131.00,21.51){\line(0,0){0.00}}
\put(131.00,21.51){\line(0,0){0.00}}
\put(131.00,21.51){\line(0,0){0.00}}
\put(131.00,21.51){\line(0,0){0.00}}
\put(131.00,21.51){\line(0,0){0.00}}
\put(110.33,9.89){\line(1,0){15.67}}
\put(136.00,9.89){\line(1,0){15.00}}
\put(151.00,34.84){\line(-1,0){14.67}}
\put(126.00,34.84){\line(-1,0){15.33}}
\put(131.00,21.51){\makebox(0,0)[cc]{F}}
\put(141.99,34.84){\line(0,-1){4.30}}
\put(141.99,27.96){\line(0,-1){3.87}}
\put(141.99,21.51){\line(0,-1){3.01}}
\put(141.99,18.92){\line(0,-1){1.29}}
\put(141.99,15.48){\line(0,-1){3.44}}
\put(141.99,11.18){\line(0,-1){1.29}}
\put(67.66,34.84){\line(0,-1){4.30}}
\put(67.66,27.96){\line(0,-1){3.87}}
\put(67.66,21.51){\line(0,-1){3.01}}
\put(67.66,18.92){\line(0,-1){1.29}}
\put(67.66,15.48){\line(0,-1){3.44}}
\put(67.66,11.18){\line(0,-1){1.29}}
\put(30.00,21.51){\line(0,0){0.00}}
\put(30.00,21.51){\line(0,0){0.00}}
\put(30.00,21.51){\line(0,0){0.00}}
\put(30.00,21.51){\line(0,0){0.00}}
\put(30.00,21.51){\line(0,0){0.00}}
\put(9.33,9.89){\line(1,0){15.67}}
\put(35.00,9.89){\line(1,0){15.00}}
\put(50.00,34.84){\line(-1,0){14.67}}
\put(25.00,34.84){\line(-1,0){15.33}}
\put(30.00,21.51){\makebox(0,0)[cc]{F}}
\put(18.33,32.26){\circle*{0.86}}
\put(18.33,27.96){\circle*{0.86}}
\put(18.33,24.09){\circle*{0.86}}
\put(18.33,19.78){\circle*{0.86}}
\put(18.33,15.48){\circle*{0.86}}
\put(18.33,11.18){\circle*{0.86}}
\put(41.00,9.46){\line(0,1){3.87}}
\put(41.00,14.62){\line(0,1){3.87}}
\put(41.00,20.21){\line(0,1){3.87}}
\put(41.00,25.80){\line(0,1){3.87}}
\put(41.00,31.39){\line(0,1){3.44}}
\put(30.00,5.16){\makebox(0,0)[cc]{j}}
\put(131.00,5.59){\makebox(0,0)[cc]{l}}
\put(131.00,40.43){\makebox(0,0)[cc]{i}}
\end{picture}
\\
\\
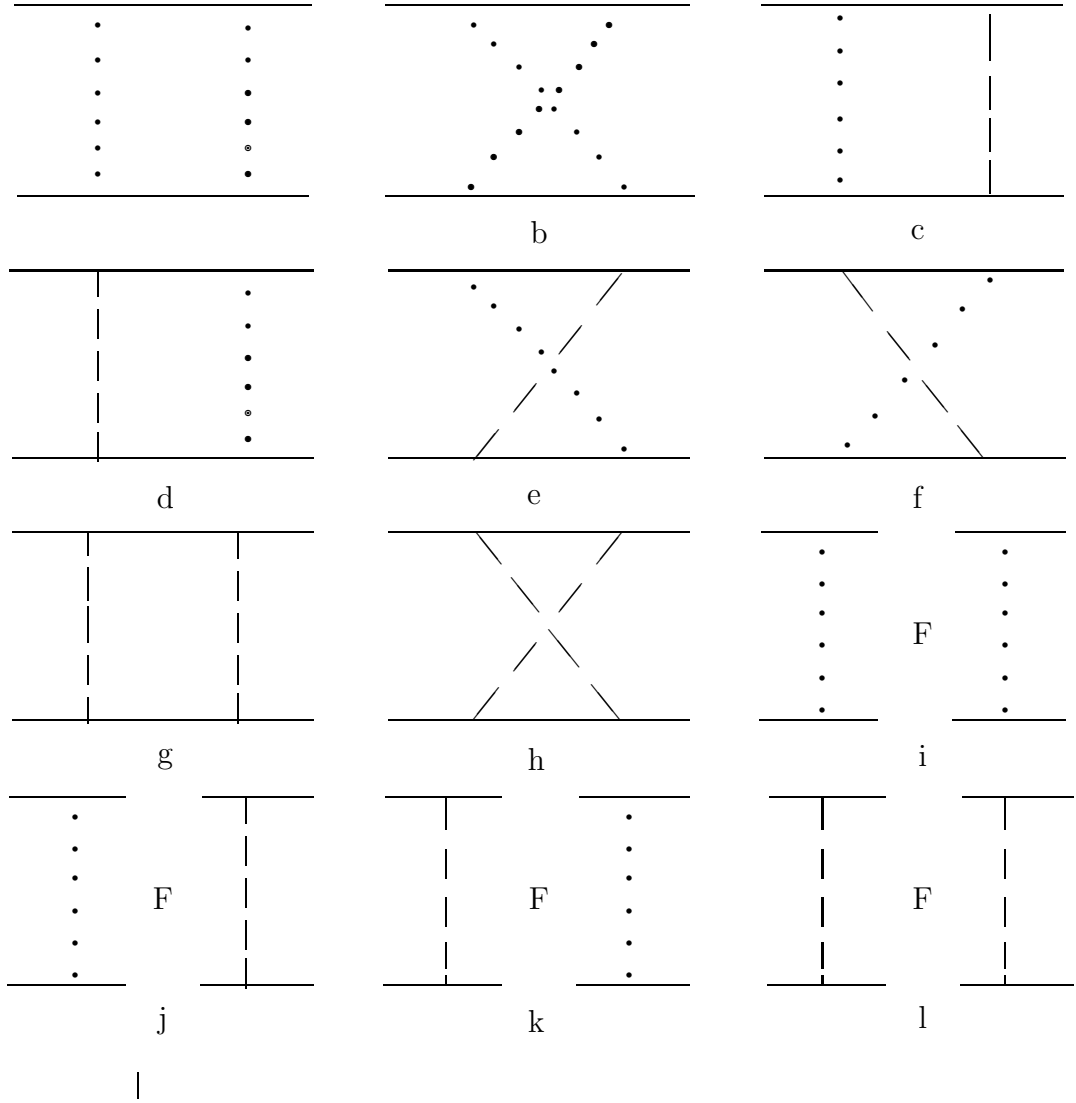
{\bf Fig. 1.} The diagrams for calculation of the quasipotential in the fourth
order of perturbation theory(unequal mass case). The dot line corresponds
to the Coulomb photon; the dashed line, to the transverse photon.

\end{document}